\documentclass[aps,twocolumn,showpacs]{revtex4}
\usepackage{graphicx}

\begin{document}

\title{Robust superconductivity and transport properties in (Li${_{1-x}}$Fe${_x}$)OHFeSe single crystals}

\author{Hai Lin, Jie Xing, Xiyu Zhu, Huan Yang and Hai-Hu Wen}\email{hhwen@nju.edu.cn}

\affiliation{Center for Superconducting Physics and Materials,
National Laboratory of Solid State Microstructures and Department
of Physics, National Center of Microstructures and Quantum
Manipulation, Nanjing University, Nanjing 210093, China}

\date{\today}

\begin{abstract}
The recently discovered (Li${_{1-x}}$Fe${_x}$)OHFeSe
superconductor with $T_c$ about 40K provides a good platform for
investigating the magnetization and electrical transport
properties of FeSe-based superconductors. By using a hydrothermal
ion-exchange method, we have successfully grown crystals of
(Li${_{1-x}}$Fe${_x}$)OHFeSe. X-ray diffraction on the sample
shows the single crystalline PbO-type structure with the c-axis
preferential orientation. Magnetic susceptibility and resistive
measurements show an onset superconducting transition at around
${T_c}$=38.3K. Using the magnetization hysteresis loops and Bean
critical state model, a large critical current ${J_s}$ is observed
in low temperature region. The critical current density is
suppressed exponentially with increasing magnetic field.
Temperature dependencies of resistivity under various currents and
fields are measured, revealing a robust superconducting current
density and bulk superconductivity.
\end{abstract}

 \maketitle

Since the discovery of high temperature superconductivity in
LaFeAsO${_{1-x}}$F${_x}$\cite{HosonoJACS2008}, at least eight
different structures of iron-based superconductors have been found
in succession\cite{ChuCW,WenHH,RGreene}. Therein, FeAs-based as
well as FeSe-based superconductors are two most common and
important families. For the electric neutrality of FeSe-layers, by
the normal high temperature sintering method, no spacer layer can
be easily intercalated in between FeSe-layers, whose structure is
similar to FeAs-layers. As far as we know, FeSe-based
superconductors cover the phases of Fe${_{1+x}}$Se, monolayer FeSe
film on SrTiO$_3$ substrate, $A$${_x}$Fe${_{2-y}}$Se$_2$,
$A_x$(NH$_3$)$_y$Fe$_{2-z}$Se$_2$ ($A$ is alkali or alkali-earth metal),
Li${_x}X_y$Fe${_{2-z}}$Se${_2}$( $X$ is an organic molecule) and so
on\cite{review1}. Among them, Fe${_{1+x}}$Se single crystal has a
low superconducting transition temperature of only about 8K at
ambient pressure\cite{FeSe}. Although monolayer FeSe film with
$T_c$ about 65K\cite{Xueqikun} indicates a possibility of high ${T_c}$ superconductivity in FeSe-based superconductors, it is made on oxide
substrates like SrTiO${_3}$ by MBE and is extremely sensitive to
air\cite{Xueqikun,LiuD}. The FeSe monolayer thin film without
covering with FeTe layers will be damaged after being taken out of
the vacuum chamber. So it is very hard to measure the
magnetization and electrical transport properties in situ. For the
$A$${_x}$Fe${_{2-y}}$Se${_2}$ system with superconducting $T_c$ at
about 32K, single crystals can be grown by the high temperature
sintering method \cite{KFeSe1,HanFei}. However, the
superconducting phase is always inter-grow with the insulating
phase $A$${_{0.8}}$Fe${_{1.6}}$Se${_2}$ (or called as the 245
phase) which has a $\sqrt{5}$$\times$$\sqrt{5}$ ordered structure
of Fe-vacancies\cite{KFeSeSepPhase}. In the phase of
$A$${_{x}}$Fe${_{2-y}}$Se${_2}$, through the scanning electron
microscopy (SEM) analysis, it was shown the the superconducting
phase takes over only about 20\% volume of the total phase. Thus,
this phase separation clearly obstacles the investigation of
intrinsic superconducting properties of the FeSe-based
superconducting phase. Meanwhile, the
$A$$_x$(NH$_3$)$_y$Fe${_{2-z}}$Se$_2$ and
Li${_x}X_y$Fe$_{2-z}$Se$_2$ phases are very sensitive to air. Up to now, no single crystals of them are available for
measurements.

The newly found superconductor (Li${_{1-x}}$Fe${_x}$)OHFeSe can be
conveniently synthesized by the hydrothermal
method\cite{Chenxianhui}. Comparing with other FeSe-based
superconductors, the superconducting transition temperature
${T_c}$ of about 40K of (Li${_{1-x}}$Fe${_x}$)OHFeSe is much
higher than that in Fe${_{1+x}}$Se and comparable to the monolayer
FeSe thin film system\cite{Xueqikun}. It was shown that the
(Li${_{1-x}}$Fe${_x}$)OHFeSe phase has very little Fe-vacancies in
the FeSe-layers with a measured ration about
Fe:Se=0.98:1\cite{Johrendt}. Therefore, this system has some
advantages over the $A$${_x}$Fe${_{2-y}}$Se${_2}$ which has a
great quantity of Fe-vacancy and accompanies with the phase
separation. Moreover, by a hydrothermal ion-exchange method, one
can obtain crystals of (Li${_{1-x}}$Fe${_x}$)OHFeSe with large
sizes and good quality\cite{Dong}. And the single crystals can
keep stable in the air for a period of time. Therefore, it is a
good platform to study the intrinsic properties of FeSe-based
superconductors, from which a further comprehension to the
mechanism of iron-based superconductors can be acquired. In the
(Li${_{1-x}}$Fe${_x}$)OHFeSe systems, several measurements have
already been employed, such as scanning tunneling microscopy
(STM), angle resolved photoemmison spectroscopy (ARPES), ionic
field gating effect, etc.\cite{WenSTM,ZhouArpes,ChenIFET}. Our
previous STM work\cite{WenSTM}, based on the high quality
crystals, has clearly indicated the presence of double anisotropic
gaps in (Li${_{1-x}}$Fe${_x}$)OHFeSe, which mimics that in the
monolayer FeSe thin film. Previously Dong et al. have done some
electrical transport measurements on the single
crystals\cite{Dong}. Here, we show some further measurements on
magnetic and electrical properties of
(Li${_{1-x}}$Fe${_x}$)OHFeSe, in order to know how robust the
superconductivity is, and whether it is bulk superconductive or it
is phase separated, like in $A$${_x}$Fe${_{2-y}}$Se${_2}$.

\begin{figure}
\includegraphics[width=8.5cm]{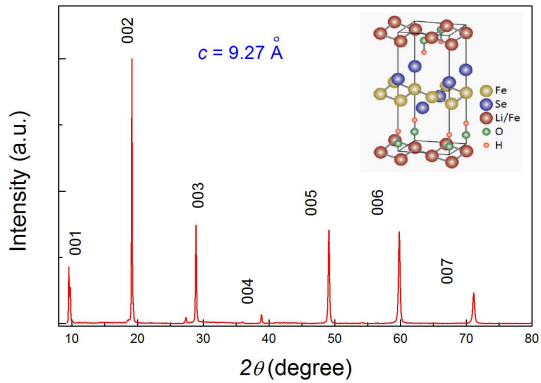}
\caption{(Color online) X-ray diffraction patterns for the
(Li${_{1-x}}$Fe${_x}$)OHFeSe crystal. One can see the predominant
(00l) indices. The inset is a schematic structure of tetragonal
(Li${_{1-x}}$Fe${_x}$)OHFeSe.} \label{fig1}
\end{figure}

The (Li${_{1-x}}$Fe${_x}$)OHFeSe crystals investigated in this
work are synthesized using a hydrothermal ion-exchange method, as
reported previously\cite{Dong}. Firstly,
K${_{0.8}}$Fe${_2}$Se${_2}$ crystals were grown using the
self-flux method. Next, LiOH (J${\&}$K, 99${\%}$ purity) liquid
was dissolved in deionized water in a teflon-linked
stainless-steel autoclave (volume 50mL). Then, iron powder
(Aladdin Industrial, 99.99${\%}$ purity), selenourea (J${\&}$K,
99.9${\%}$ purity), and several pieces of
K${_{0.8}}$Fe${_2}$Se${_2}$ crystals were added to the solution.
After that, the autoclave was sealed and heated up to 120
$^{\circ}$C followed by staying for 40 to 50 hours. Finally, the
(Li${_{1-x}}$Fe${_x}$)OHFeSe crystals with a metallic-grey colored surface can be obtained by leaching. The X-ray diffraction (XRD)
measurements were performed on a Bruker D8 Advanced diffractometer
with the Cu-K$_\alpha$ radiation. DC magnetization measurements
were carried out with a Quantum Design instrument SQUID-VSM-7T.
The resistive measurements were done with the standard four-probe
method on a Quantum Design instrument Physical Property
Measurement System (PPMS). In measuring the resistivity, the
current was switched from positive to negative alternatively in
order to remove the contacting thermal power.

\begin{figure}
\includegraphics[width=8.2cm]{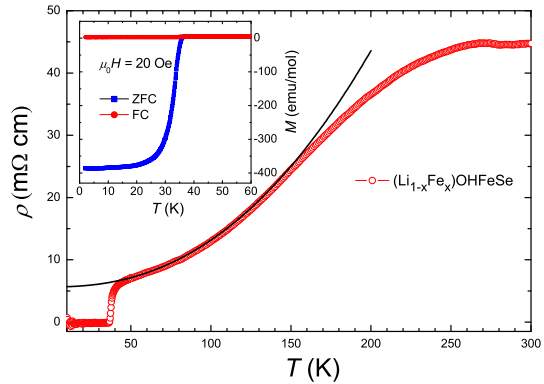}
\caption{(Color online) Temperature dependence of resistivity for
the (Li${_{1-x}}$Fe${_x}$)OHFeSe crystal at zero field with
measuring current of 20$\mu$A. The solid line shows the fit in the
low temperature range by the formula $\rho(0)$+A$T$${^n}$. The
upper inset is the temperature dependence of magnetic
susceptibility measured in both ZFC and FC modes for a sample
taken from the same batch, with an applied field of 20Oe parallel
to c-axis.} \label{fig2}
\end{figure}

Fig.~\ref{fig1} shows the X-ray diffraction (XRD) spectra for the
(Li${_{1-x}}$Fe${_x}$)OHFeSe single crystal. The sample turns out
to be very layered feature and can be easily cleaved. Only (00l)
reflections can be seen in the XRD pattern, indicating highly
orientation along the c-axis. The $l$ containing both odd and even
numbers, which indicates the structural changing from I4/mmm of
K${_{0.8}}$Fe${_2}$Se${_2}$ to P4/nmm of
(Li${_{1-x}}$Fe${_x}$)OHFeSe. The c-axis lattice constant is about
9.27$\AA$ which is close to the reported results in
(Li${_{1-x}}$Fe${_x}$)OHFeSe\cite{Chenxianhui}. The inset shows
the schematic structure of (Li${_{1-x}}$Fe${_x}$)OHFeSe which has a
typical layered structure of iron-based superconductors.

In Fig.~\ref{fig2}, we present the temperature dependence of
resistivity from 10K to 300K at zero field with a measuring
current of 20$\mu$A. The resistivity decreases monotonically from
room temperature to the lower temperature, which shows a highly
metallic conductivity. The solid line shows the fit in the low
temperature range by $\rho(T)$=$\rho(0)$+A$T$${^n}$ with
$\rho(0)$= 5.65m${\Omega}$*cm , A= 0.00016m${\Omega}$*cm/K${^2}$,
n= 2.32. The residual resistivity ratio, defined as
RRR=$\rho(300K)/\rho(0K)\approx $7.9, which is comparable to the
previous reported value\cite{Dong}. In the normal state, the
temperature dependent resistivity here exhibits a positive
curvature and the ratio d$\rho$/d$T$ becomes smaller at high
temperature. This is similar to Fe${_{1+x}}$Se single
crystals\cite{FeSe}, but quite different from some
$A$${_x}$Fe${_{2-y}}$Se${_2}$ single crystals\cite{KFeSe1}, where
a positive curvature at low temperature and a negative curvature
at high temperature are generally observed. In the low temperature
region, an abrupt resistivity drop can be clearly seen. Using the
criterion of 90${\%}$ of the normal state resistivity ${\rho_n}$,
the onset superconducting transition temperature ${T_c^{onset}}$
is determined, which gives a value of about 38.3K. The inset
presents the temperature dependence of magnetic susceptibility
under an applied field of 20Oe measured in the zero-field-cooled
(ZFC) and field-cooled (FC) modes. The sample for the
magnetization measurements is one taken from the same batch of
that for the resistive measurements. Similar to the resistivity
curves, a sharp superconducting transition can be clearly seen at
about 38K in the magnetic measurements. The transition temperature
is higher than that in $A$${_x}$Fe${_{2-y}}$Se${_2}$($A$ = K, Rb,
Cs, Tl)\cite{KFeSe, RbFeSe, CsFeSe, TlFeSe} and Fe${_{1+x}}$Se at
ambient pressure\cite{FeSe}. Furthermore, the magnetic
susceptibility measured in the ZFC mode reveals an almost fully
superconducting screening effect.

\begin{figure}
\includegraphics[width=8.5cm]{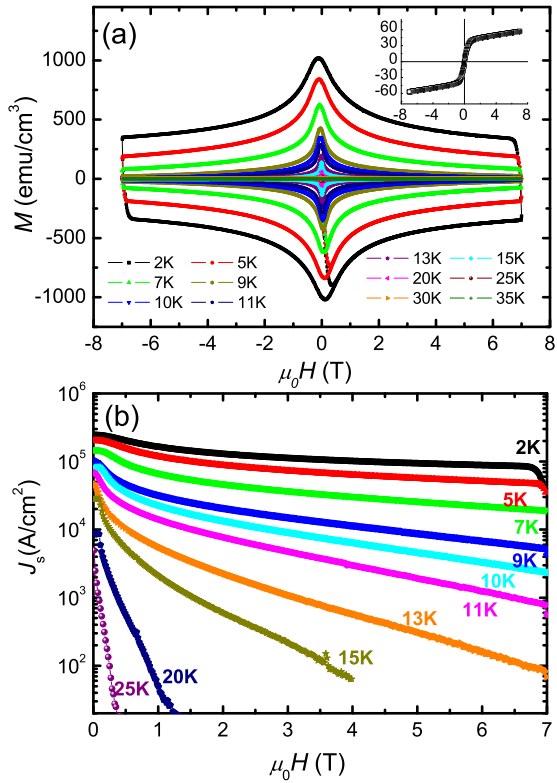}
\caption{(Color online)(a) Magnetization hysteresis loops of the
(Li${_{1-x}}$Fe${_x}$)OHFeSe crystal at various temperatures below
$T_c$, which have deducted the ferromagnetic background measured
at 50K as shown in the inset. (b) Magnetic field dependence of the calculated
superconducting current density based on the Bean critical state
model at temperatures ranging from 2K to 25K.} \label{fig3}
\end{figure}

Fig.~\ref{fig3}(a) presents the magnetization hysteresis
loops (MHLs) of the (Li${_{1-x}}$Fe${_x}$)OHFeSe crystals at
various temperatures below ${T_c}$. The MHLs presented here have
been deducted a weak ferromagnetic background signal from the raw data
measured at 50K. The background is shown in the inset of
Fig.~\ref{fig3}(a). This weak ferromagnetic signal may be induced
by some impurities, or it is an intrinsic property of the system
due to the substitution of Fe to the Li atoms\cite{Johrendt}. This
needs to be further resolved by more investigations. During the
measurements, the magnetic field is always perpendicular to the ab
plane of the sample. Here, we determine the width $\Delta$$M$ of MHLs, where
$\Delta$$M$ is $M^{down}$-${M^{up}}$. $M^{down}$(${M^{up}}$) is
the magnetization at a certain magnetic field in the decreasing
(increasing)-field process. It is interesting to note that the
maximum of the MHL width $\Delta$$M$$\approx $ 2000 emu/cm${^3}$
observed at 2K is comparable to other bulk iron-based
superconductors, such as the optimally doped
BaFe${_{2-x}}$Co${_{x}}$As${_2}$. This value is at least one order
of magnitude larger than that of
K${_x}$Fe${_{2-y}}$Se${_2}$\cite{KFeSe}, which reveals the good
quality and bulk pinning of our samples. The monotonically
decreasing of ${M}$ with increasing field reveals the absence of
fish-tail effect below 7T. For a further analysis, we calculate
the critical current density using the Bean critical state
model\cite{Bean}. In this model, the superconducting critical
current density ${J_s}$ is expressed by

\begin{equation}
J_s=20\frac{\Delta M}{a(1-a/3b)}
\end{equation}

where a(cm) and b(cm) (a${\leq}$b) are the in-plane sample sizes.
The calculated results are illustrated in Fig.~\ref{fig3}(b) in a
semi-logarithmic scale. Obviously, the critical current density
$J_s$ of (Li${_{1-x}}$Fe${_x}$)OHFeSe is weakly dependent with the
magnetic field at temperatures below 9K. The critical current
density $J_s$ is as large as 2.47$\times$10${^5}$A/cm${^2}$ at 2K in
the low field limit. The magnitude is comparable to that in the
optimally doped BaFe${_{2-x}}$Co${_{x}}$As${_2}$, and much larger than
10${^4}$A/cm${^2}$ in the Fe${_{1+x}}$Se and
K${_x}$Fe${_{2-y}}$Se${_2}$\cite{FeSeJs,KFeSeJs1,KFeSeJs2}
crystals. At higher temperatures over 9K, ${J_s}$ decreases
rapidly with increasing temperature and finally becomes
vanishingly small after 25K. In addition, the ${J_s}$ in the high
temperatures region is depressed exponentially with the increasing
field. And it is interesting that this type of MHL shape looks
very similar to those in many superconductors, but is very
different from K${_x}$Fe${_{2-y}}$Se${_2}$\cite{KFeSeJs2} in which
phase separations are expected.

\begin{figure}
\includegraphics[width=8.5cm]{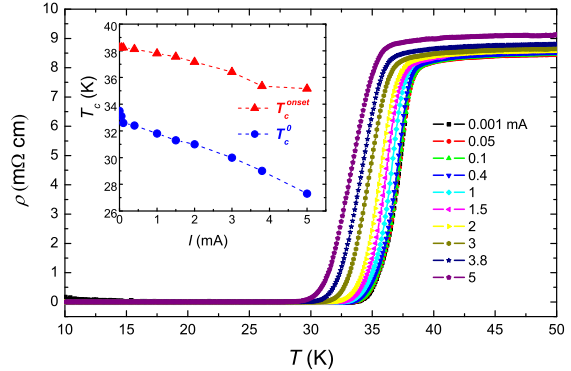}
\caption{(Color online) (a) Temperature dependence of resistivity
for the (Li${_{1-x}}$Fe${_x}$)OHFeSe crystal measured with various
currents which are applied always in the ab-plane. The inset shows
the current dependence of ${T_c}$, where ${T_c^{onset}}$ is
determined with a criterion of 90${\%}$${\rho_n}$, and the zero
transition temperature ${T_c^0}$ was determined with
1${\%}$${\rho_n}$.} \label{fig4}
\end{figure}

In order to check the homogeneity of superconductivity in
(Li${_{1-x}}$Fe${_x}$)OHFeSe crystals, we measured the temperature
dependence of resistivity at various currents from 10K to 50K, and
the data are presented in Fig.~\ref{fig4}. The current ${I}$ is
always applied in the ab-plane. It is found that when a larger
measuring current is used, the normal state resistivity ${\rho_n}$
becomes slightly increased. The enhancement of the normal state
resistivity may be induced by the local heating effect when the
measuring current is high. It is interesting to note that the
value of ${T_c}$ drops down slowly when the measuring current
${I}$ is increased from 0.001mA to 5mA (corresponding to the current
density of about 0.0136 to 68A/cm$^2$). The current dependence of
the superconducting transition temperature $T_c$ is shown in the
inset of Fig.~\ref{fig4}. Here, ${T_c^{onset}}$ is determined with
the criterion of 90${\%}$${\rho_n}$ and ${T_c^0}$ with
1${\%}$${\rho_n}$. This weak suppression of $T_c$ by the measuring
current density may suggest that the sample should not have the
phase separation like that in K${_x}$Fe${_{2-y}}$Se${_2}$. If
phase separation would exist in the system, the superconducting
transition and the related transition temperature should be
strongly influenced by the measuring current density. This seems
not the case here. This conclusion is also consistent with the
large MHL width and $J_c$ value observed in the
(Li${_{1-x}}$Fe${_x}$)OHFeSe crystal.

\begin{figure}
\includegraphics[width=8.5cm]{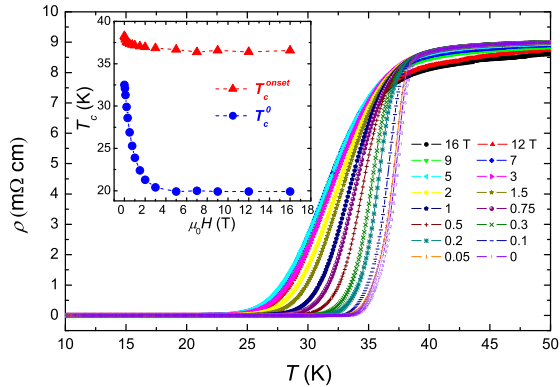}
\caption{(Color online) Temperature dependence of resistivity with
the magnetic field applied parallel to c-axis. The inset shows the
field dependence of ${T_c}$. ${T_c^{onset}}$ is determined with
criterion of 90${\%}$${\rho_n}$ and ${T_c^0}$ with
1${\%}$${\rho_n}$.} \label{fig5}
\end{figure}

In Fig.~\ref{fig5}, we show the temperature dependence of
resistivity for the (Li${_{1-x}}$Fe${_x}$)OHFeSe crystal at zero
field and various magnetic fields with the field directions
parallel to c-axis, while the current is always applied in the
ab-plane. As the applied field is increased, the superconducting
transition temperature is suppressed gradually, and the normal state
resistivity ${\rho_n}$ reveals a negative magnetoresistivity. This
abnormal negative magnetoresistivity has not been found in the
previous report\cite{Dong}, and might be caused by some magnetic
impurities arising from the partial substitution of Li by Fe in
the Li layers. A negative magnetoresistance is expected in a
system with magnetic scattering centers. In addition, the magnetic
field induces a clear broadening of the superconducting
transition. This broadened transition is induced by the vortex
motion since the (Li${_{1-x}}$Fe${_x}$)OHFeSe system has a large
spatial distance between the FeSe layers and thus the anisotropy
is quite high. However, the upper critical field $H_{c2}$ defined
by 90\%$\rho_n$ remains robust, suggesting a strong pairing
strength since $H_{c2}$ is proportional to $\Delta_s^2$ with
$\Delta_s$ the superconducting gap. Through our measurements of
the magnetization and the resistive transitions under different
magnetic fields and current, and the thoughtful analysis, we can
conclude that the superconductivity is robust, uniform without the
phase separation as that occurring in K${_x}$Fe${_{2-y}}$Se${_2}$.

In summary, we successfully synthesized the
(Li${_{1-x}}$Fe${_x}$)OHFeSe crystals with large size and good
quality by the hydrothermal ion-exchange method. The magnetic
hysteresis loops at various temperatures are measured which
exhibit a symmetric shape, indicating a vortex bulk pinning
property and thus bulk superconductivity. By using the Bean
critical state model, we calculated the superconducting current
density ${J_s}$ which amounts to 2.47$\times$10$^5$ A/cm$^2$ at 2K
and zero magnetic field. This value of ${J_s}$ of
(Li${_{1-x}}$Fe${_x}$)OHFeSe is very large compared with that in
K${_x}$Fe${_{2-y}}$Se${_2}$. The latter was proved to have phase
separation. In addition, we measured the temperature dependence of
resistivity with different transport currents in
(Li${_{1-x}}$Fe${_x}$)OHFeSe. It is found that the superconducting
transition temperature drops down slightly when the measuring current
density is increased from 0.0136 to 68A/cm${^2}$, suggesting again the
absence of phase separation. Finally the resistive transition has
been measured under different magnetic fields. A clear broadening
of the transition is observed, which indicates a strong vortex
motion due to the high anisotropy of the system.

\section*{ACKNOWLEDGMENTS}
This work was supported by the National Natural Science Foundation
of China(Grant No.11534005), the Ministry of Science and
Technology of China (Grant Nos.2011CBA00102, 2012CB821403).

\end{document}